\documentclass[twocolumn,showpacs,preprintnumbers]{revtex4}
\usepackage{graphicx}
\usepackage{dcolumn}
\usepackage{bm}

\begin{document}

\title{Nonergodic Brownian Dynamics and the Fluctuation-Dissipation Theorem }
\author{Jing-Dong Bao,$^{1}$ Yi-Zhong Zhuo,$^2$ Fernando A. Oliveira,$^3$ and
Peter H\"{a}nggi$^4$}
\affiliation{$^1$Department of Physics, Beijing Normal University, Beijing 100875, China\\
$^2$China Institute of Atomic Energy, P. O. Box 275, Beijing 102413, China\\
$^3$Institute of Physics and International Center of Condensed Matter
Physics, University of Brasilia-CP 04513, 70919-970, Brasilia-DF, Brazil\\
$^4$Institute of Physics, University of Augsburg, D-86135
Augsburg, Germany}
\date{\today }

\begin{abstract}
Nonergodic Brownian motion  is elucidated within the framework of the
generalized Langevin equation. For thermal noise yielding either
a vanishing or a divergent zero-frequency friction strength, the non-Markovian Browninan dynamics
exhibits a riveting, anomalous diffusion
behavior being characterized by a ballistic or possibly also a localized dynamics. As
a consequence, such tailored thermal noise may cause a net
acceleration of directed transport in a rocking Brownian motor. Two
notable conditions  for the thermal noise are identified in order to guarantee the
fluctuation-dissipation theorem of first kind.
\end{abstract}

 \maketitle

The phenomenon of Brownian motion has assumed a fundamental and influential role in the
development of  thermodynamical and statistical theories and
continues to do so as an inspiring source for active research in
various fields of natural sciences \cite{HM2005}. The  Brownian
dynamics can conveniently be described by a generalized Langevin equation (GLE), reading:
$m\dot{v}(t)+m\int_{0}^{t}\gamma (t,t^{\prime })v(t^{\prime
})dt^{\prime }+\partial _{x}U(x,t)=\varepsilon (t)$. The thermal
random force $ \varepsilon (t)$ is assumed to be uncorrelated with
the initial velocity. The memory friction  $\gamma (t,t^{\prime })$
 is typically related to the correlation of the random forces \cite{kubo,toda}. Kubo \cite{kubo}
 has addressed the  common behavior of
a classical equilibrium bath by setting (i) $\langle \varepsilon
(t)\varepsilon (t^{\prime })\rangle =mk_{B}T\gamma (t,t^{\prime})$
with (ii) $\gamma (t,t^{\prime })=\gamma (|t-t^{\prime}|)$ being time-homogeneous.
The one-sided Fourier transform consequently then
obeys  $Re[\tilde{\gamma}(\omega )]\geq 0$,
for real-valued $\omega$. Here $k_{B}$ is the Boltzmann constant and
$T$ denotes the bath temperature. The validity of the GLE is thus
restricted to the case of a stationary thermal noise obeying the
above two conditions; e.g. see in Refs. \cite{kubo, toda,pot}.  The complex-valued mobility
$\tilde{\mu}(\omega)$ and the
memory-friction $\tilde{\gamma}(\omega)$, respectively, are given in terms of the
Fourier-Laplace transforms of the time-homogeneous correlations of the particle velocity $v(t)$
and the environmental noise $\epsilon(t)$, respectively, as
\begin{eqnarray}
 \tilde{\mu}(\omega )
=\frac{1}{k_{B}T}\int_{0}^{\infty } dt \exp(-i\omega t) \langle
v(t_{0})v(t_{0}+t)\rangle,\\
 m\tilde{\gamma}(\omega)=\frac{1}{k_{B}T}\int_{0}^{\infty } dt \exp(-i\omega t)
 \langle \varepsilon (t_{0})\varepsilon
(t_{0}+t)\rangle.
\end{eqnarray}
To differentiate between these two relations (1) and (2), one refers
to the first  relation as the FDT of the {\it first kind} and the
second one as  the FDT of the {\it second kind}, respectively
\cite{kubo,toda}. The former characterizes the relaxation of the
average particle velocity, while the latter yields the generalized
susceptibility, describing the response of the bath degrees of
freedom.

These two  FDT relations play a central role in the theory of thermal Brownian
motion; its validation and generalization presents a timely
subject that is presently hotly debated, both within theory and experiment.

With this work we aim at demonstrating that  Kubo's
conditions for the equilibrium bath are generally not complete within the
framework of linear response theory. This is so, because the
existence of anomalous diffusion has not been considered in the
original treatment by Kubo and others. We thus not only seek to complete set of conditions
for general  Brownian motion to obey an ergodic behavior but, more importantly, we like to dwell on
rich behavior that emerges if ergodicity does not hold. Throughout the following we use the
nomenclature of {\it ergodicity} in the context of the theory of stochastic processes,
i.e. for {\it open} systems that are
subjected to noise: Then the process is ergodic if the temporal long-time
average of a state function equals its
stationary ensemble average \cite{papoulis}. Thus, we work beyond the well-known case of
isolated systems obeying an area preserving, Hamiltonian dynamics \cite{toda,VK}.
Put differently, our main objective is to extend the theory of classical Brownian motion
by focussing  on the intricacies of a possible
non-ergodic, non-Markovian dynamics \cite{mor,costa,lee,lutz,mok,rubi}.
In particular, the asymptotic long-time
statistical probability density will be termed ergodic if it approaches a
stationary state that is independent of the choice of the initial preparation.

\textit{Noisy dynamics exhibiting nonergodic behavior}. As we shall
show below, systems possessing via the FDT of the second kind no
finite friction at zero frequency typically exhibit an non-ergodic
behavior. Thus, there is an abundance of open system dynamics for
which ergodicity is not granted. The thermal fluctuations result
typically from a coupling to environmental degrees of freedom. In
the cases of bi-linear system-bath coupling these are characterized
by a corresponding spectral density of  bath modes, $J(\omega)$,
obeying $Re[\tilde{\gamma}(\omega )] = J(\omega)/ m\omega$
\cite{HTB,gra87,RI,chen}. Thus, if the coupling to low frequency
modes is weak, as it is the case with optical-like phonons
\cite{mor, RI}, or broadband colored noise \cite{bao2003},  or also
for the  celebrated case with a blackbody radiation field of the
Rayleigh-Jeans type \cite{for} the static friction  vanishes at zero
frequency $\omega=0$. Then, no  efficient mechanism necessary for
the compliance of the  ergodic behavior is at work. Other situations
that come to mind involve the vortex diffusion in magnetic fields
\cite{ao}, or  diverse other open dynamics with an inherent
velocity-dependent system-bath coupling \cite{for,Bai2005,bao06}.
Note also that a typical solid state Drude bath spectrum is
proportional to $\omega^3$, thus yielding as well a vanishing,
zero-frequency friction \cite{gra87,RI}.

\textit{Conditions for the validity of the FDT}. The GLE was originally
derived by Mori by use of the Gram-Schmidt procedure; it was re-obtained
by Lee using the recurrence relations method \cite{lee2}. Starting out from the well-known
system-plus-oscillator-reservoir model detailed in Ref. \cite{zwanzig},
the equations of motions directly yield the
GLE with a stationary, correlated thermal noise $\epsilon(t)$. The
initial coordinate and velocity of each oscillator are assumed
to be distributed according to thermal equilibrium.
We now add two additional mixing conditions to the above given, well-known
relations (i) and (ii). These read: (iii) $\lim_{s\rightarrow
0}[s^{-1}\hat{\gamma}(s)]\rightarrow \infty $ and (iv)
$\lim_{s\rightarrow 0}[s\hat{\gamma}(s)]=0$, where $\hat{\gamma}(s)$
is the Laplace transform of the memory friction kernel that
enters the GLE. Our terminology will be as follows: If
one of the above given four conditions (i)-(iv) is {\it not}
satisfied we term the bath -- in the specified order (i) -- (iv):
(i) a {\it nonequilibrium bath} \cite{grabert80}, (ii) a {\it relaxing}
 (or ageing) one \cite{pot,abo},  (iii) a {\it ballistic} one
\cite{mor,bao2003}, or (iv) a {\it nonergodic-bath}, respectively.
The notion embracing the latter two cases (iii-iv) will also  be referred to as
{\it weak nonequilibrium heat baths}.

For the force-free GLE, i.e. with $U(x)=0$, the  two-time
velocity correlation function (VCF) reads,
\begin{eqnarray}
C_{vv}(t_{1},t_{2})
&=&\{v^2(0)\}b^2+\frac{k_{B}T}{m}\left[\frac{S(\tau)}{k_BT/m}+b-b^2\right]\nonumber \\
&&+\left( \{v^{2}(0)\}-\frac{k_{B}T}{m}\right) A(t_{1},t_{2}),
\end{eqnarray}
where $\tau=|t_1-t_2|$ and the explicit forms of the two relaxation functions $A$ and $S$
are detailed in Ref. \cite{Bai2005}. Note that depending on the choice of initial preparation
this correlation generally is {\it not} time-homogeneous. Herein, we indicate by $\{\cdots \}$ the
average with respect to the initial preparation of the state variables.
The relevant, generally non-vanishing quantity $b$ is given by:
\begin{equation}
b=[1+\lim_{s\rightarrow 0}(\hat{\gamma}(s)/s)]^{-1}\;.
\end{equation}

The non-equilibrium, time-inhomogeneous relaxation part $A(t_{1},t_{2})$
obeys $A(t_1\to\infty, t_2\to\infty)=0$. The asymptotic
equal-time dynamics is found to read \cite{Bai2005}: $\{\langle
v^2\rangle\}_{st}=\{v^2(0)\}b^2+k_{B}Tm^{-1}(1-b^2)$. Moreover, the
asymptotic velocity correlation function emerges as
 $C_{vv,\textmd{st}}=bk_{B}T/m$  for the case that we set the
initial velocity variance in accordance with the thermal equilibrium
value  $ \{v^2(0)\}=k_{B}T/m$. These results follow for $b\neq 0$,
thus implying that the condition (iii) is not satisfied (case with a
ballistic bath). This in turn requires that
$\hat{\gamma}(0)=\int^{\infty}_0\gamma(t)dt=0$, meaning a vanishing
effective friction at zero frequency,  or equivalently a vanishing
of the spectral density of the  noise $\varepsilon(t)$ at
$\omega=0$, see in (2) \cite{costa,bao2003}. This causes a breakdown
of the FDT of the first kind because of the preparation-dependent,
i.e. $v(0)$-dependent asymptotic result. This is so although  the
FDT of the second kind in (2) is valid.

Next, we discuss the situation with a nonergodic bath with
$\lim_{s\rightarrow 0}[s\hat{\gamma}(s)]\neq 0$. Towards this goal,
we  consider the solution $v(t)$ of the GLE for a case where $b\neq
0$ constitutes a stationary noise source that drives the force-free
Brownian GLE-dynamics. Put differently,  we set for the acting
thermal (tailored) noise source
\begin{equation}
\varepsilon_1 (t)\equiv cv(t)=c\left[ \{v(0)\}R(t)+\frac{1}{m}
\int_{0}^{t}R(t-t^{\prime })\varepsilon (t^{\prime })dt^{\prime
}\right],
\end{equation}
where the parameter $c$ denotes a coupling strength. The response
function is $R(t)=b+\sum_{j}res[\hat{R}(s_{j})]\exp (s_{j}t)$,
with its  Laplace transform reading
$\hat{R}(s)=(s+\hat{\gamma}(s))^{-1}$. The set $\{s_{j}\}$ denote the
nonzero roots of the equation $s+\hat{\gamma}(s)=0$, and the
probability density for  $v(0)$ is chosen as a Gaussian with
zero-mean and variance $\{v^{2}(0)\}=k_{B}T/m$.
In this case, the resulting memory-friction kernel
corresponding to the effective thermal noise $\varepsilon_1(t)$
is given by $\gamma _{1}(t)=c^{2}\{\langle v(t)v(0)\rangle
\}/(k_{B}T)$, yielding $\hat{\gamma}_{1}(0)\to\infty$, but
$\lim_{s\rightarrow 0}[s\hat{\gamma}_{1}(s)]=c^{2}b=$ finite. This
finding implies that the Brownian particle  with $b\neq 0$ is
immersed in a nonergodic bath.

Substituting  (5) into the GLE, we obtain an additional quadric
potential with the frequency $c^2b$ and a friction kernel given by
$\gamma_{1}(t)=\gamma(t)-c^2b$. This leads to
$\tilde{\gamma}(0)\to\infty$ so that $\tilde{\mu}(0)=0$. Therefore, the
diffusion constant vanishes, meaning  that  the motion of a
force-free particle remains bounded. This phenomenon of localization
at finite temperature  may occur for a particle dynamics that is coupled to  a set of
frozen environmental oscillators corresponding to the sub-Ohmic bath density $J(\omega)$,
being proportional to
$\omega^{\delta}$   in the limit of $\delta=0$ \cite{gra87}. A realization  provides a
harmonic chain on a Bethe lattice possessing no diffusion thus acting as a
frozen bath in this context \cite{MHLEE}.

\textit{The case with an inherent ballistic bath.} We will concentrate on  nonergodic
Brownian motion where the Kubo FDT of the second kind is assumed
to be satisfied, the condition (iii), however, is not met. A
colored noise that induces ballistic diffusion is harmonic velocity
noise (HVN) $\varepsilon(t)$ \cite{bao2005}.  The HVN obeys the
Langevin equations: $\dot{y}=\varepsilon$,
$\dot{\varepsilon}=-\Gamma \varepsilon-\Omega ^{2}y+\xi(t)$, where
$\xi(t)$ is  Gaussian white noise of vanishing mean with $\langle
\xi (t)\xi (t^{\prime })\rangle =2\eta\Gamma ^{2}k_{B}T\delta
(t-t^{\prime})$, $\eta $ is the damping coefficient, $\Gamma$ and
$\Omega $ denote the damping and frequency parameters. The second
moments of $y(0)$ and $\varepsilon(0)$ obey $ \{y^{2}(0)\}=\eta
\Gamma \Omega ^{-2}k_{B}T$ and $\{\varepsilon^{2}(0)\}=\eta \Gamma
k_{B}T$. The Laplace and  Fourier transformations of the damping
kernel function read $\hat{\gamma}(s)=\eta\Gamma s/(s^2+\Gamma
s+\Omega^2)$ and
$\textmd{Re}[\tilde{\gamma}(\omega)]=\eta\Gamma^2\omega
^{2}/[(\Omega ^{2}-\omega ^{2})^{2}+\Gamma ^{2}\omega ^{2}]$,
respectively. The latter corresponds to the spectrum of HVN which now
{\it vanishes} at zero-frequency. In presence of the FDT of the second
kind with the conditions (i) and (ii)  satisfied, the complex-valued
mobility now emerges as
\begin{eqnarray}
\tilde{\mu}(\omega)&=&\frac{b}{m}\tilde{\delta}(\omega
)+\frac{\eta
\Gamma }{m(\Omega ^{2}+\eta \Gamma )}\nonumber\\
&&\cdot\frac{\Gamma (\Omega ^{2}+\eta \Gamma )+i\omega (\Omega
^{2}+\eta \Gamma -\Gamma ^{2}-\omega ^{2})}{(\Omega ^{2}+\eta
\Gamma -\omega ^{2})^{2}+\Gamma ^{2}\omega ^{2}},
\end{eqnarray}
where $b=(1+\eta \Gamma \Omega ^{-2})^{-1}$ and
$\tilde{\delta}(\omega )=\lim_{t\rightarrow \infty }(1-\exp
(-i\omega t))/(i\omega )$. Therefore, $\tilde{ \mu}(0)\rightarrow
\infty $, yielding a diverging diffusion constant $D=\tilde{
\mu}(0)k_{B}T\rightarrow \infty $.

This free non-Markovian Brownian motion can also be recast as an
embedded, higher-dimensional Markovian process. The Fokker-Planck
equation (FPE) for the  probability density
$P(x,v,w,u,y,\varepsilon;t)$ corresponding to the set of Markovian
LEs via  introducing the appropriate set of auxiliary-variables
$(w,u,y,\varepsilon)$ obeys $\partial_{t}P-L_{FP}P=0$, where
$L_{FP}$ denotes the associated
 FPE operator:
\begin{eqnarray}
L_{FP}&=& -v\frac{\partial}{\partial
x}-m^{-1}w\frac{\partial}{\partial v}\nonumber\\
&&+(\Gamma w+\eta\Gamma v+\Omega^2y+u)\frac{\partial}{\partial
w}\nonumber\\
&&-\Omega^2(w-\varepsilon)\frac{\partial}{\partial
u}+\varepsilon\frac{\partial}{\partial y}\nonumber\\
&&+(\Gamma \varepsilon+\Omega^2 y)\frac{\partial}{\partial
\varepsilon}+\eta\Gamma^2
k_BT\frac{\partial^2}{\partial w^2}\nonumber\\
&&+\eta\Gamma^2 k_BT\frac{\partial^2}{\partial \varepsilon^2}.
\end{eqnarray}

\begin{figure}[tbp]
\includegraphics[scale=0.7]{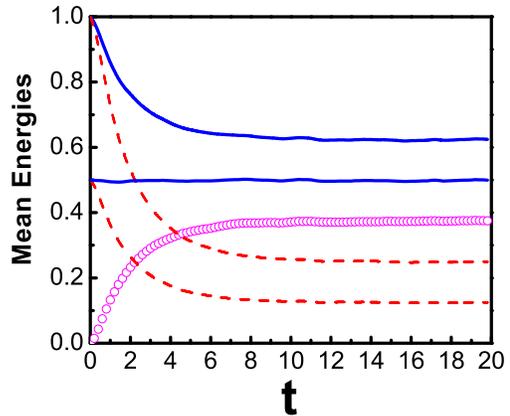}
\caption{(color online). Behavior of diverse mean energies (see text) for a nonergodic,
force-free Brownian particle {\it vs.} time  $t$. The parameters used are $m=1.0$,
$k_BT=1.0$, $\eta=0.2$, $\Gamma=5.0$, $\Omega=1.0$ $x(0)=0$, and (i)
$v(0)=\sqrt{2}$ and (ii) $1.0$, respectively, for the {\it total} energies given by the solid
lines and the {\it remnant} energies by the dashed lines, from top to
bottom. The open circles is the adsorbed energy from the bath for all
cases and also the total energy for the $v(0)=0.0$ case.} \label{1}
\end{figure}

The mean total energy of the thermal HVN-driven force-free particle
reads $\{\langle E(t)\rangle\}=\frac{1}{2}m\{\langle
v^2(t)\rangle\}=\frac{1}{2}m[\{\langle
v(t)\rangle^2\}+\{\langle(v(t)-\langle v(t)\rangle)^2\rangle\}]$.
Their stationary values are given by  first two terms in the r.h.s
of Eq. (3). In particular, the first part describes the remnant
initial kinetic energy of the particle, which is dissipated partly
by the present bath because of $\{\langle
v(t\to\infty)\rangle\}=bv(0)$. This part  vanishes for the common
case with $b=0$. The second part denotes the energy provided from
the heat bath and is independent of the initial particle velocity;
it does not relax, however, towards equilibrium. The absorbed power
of the particle is $P_{\textmd{abs}}=\eta
k_{B}T[1+\eta/(2\Gamma)+2\Omega^2/(\eta\Gamma)]^{-1}$, being smaller
than the equilibrium value $\eta k_{B}T$. This implies that the
Brownian particle is hindered in re-gaining the energy from a
thermal noise that lacks a finite, zero-frequency spectral weight.
Our numerical results are depicted in the figure 1, which are
obtained from simulating a set of Markovian LEs equivalent to the
FPE of (7). The numerics fully corroborate our theoretical findings.

\textit{Non-steady transport in ratchet potential}. A most intriguing situation refers to
Brownian motors \cite{BM}  when driven by non-ergodic Brownian motion.
Take the case of thermal
HVN  driving a Brownian particle in a rocking ratchet potential:
$U(x,t)=U_0\left[\sin(2\pi x)+c_1\sin(4\pi x)+c_2\sin(6\pi
x)\right]+A(t)x$ with $U_0=0.461$, $c_1=0.245$, and $c_2=0.04$
\cite{mac}. Here $A(t)$ is a square-wave periodic driving force that
switches forth and back between $A(t)=A_{0}$ when $2nt_{p}\leq t<(2n+1)t_{p}$ and
$A(t)=-A_{0}$ when $(2n+1)t_{p}\leq t<2(n+1)t_{p}$.
We next study whether a stationary non-equilibrium current results
that can be put to a constructive use in order to direct, separate or
shuttle particles efficiently \cite{BM}. We thus research
the resulting, time- and ensemble-averaged current with the
condition (iii)  not obeyed.

\begin{figure}[tbp]
\includegraphics[scale=0.82]{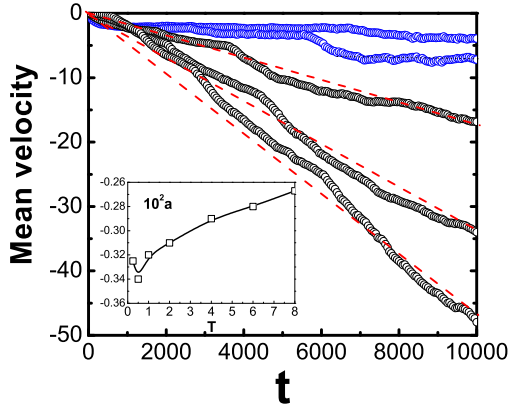}
\caption{(color online). Accelerated, (time and ensemble)-
averaged Brownian motor velocity  in a rocking ratchet  that is
driven by HVN. The used parameters are $m=1.0$, $k_BT=0.5$,
$\protect\eta =5.0$, $\Gamma=22.0$, $\Omega^2=40.0$, $t_p=25.0$,
$A_0=2.0$, $4.0$, $6.0 $, $10.0$, and $15.0$, from top to bottom.
The particles undergo a finite acceleration  $a$ (times 100) {\it vs.}
temperature $T$, being depicted with  the inset  for $A_0=10.0$.}
\label{1}
\end{figure}

Figure 2 depicts the  time-dependent, accelerated mean velocity
[$\overline{\{\langle v(t)\rangle\}}
=(2t_{p})^{-1}\int_{t}^{t+2t_{p}} dt^{\prime}\{\langle v(t^{\prime
})\rangle\} $] for various strengths of the driving force as
obtained via the simulation of the Markovian LEs corresponding to
the FPE of (7), being supplemented by the ratchet  forcing
$-U^{\prime}(x,t)$. A startling finding is that the averaged
velocity increases {\it linearly} with time. This directed
acceleration is presented by the slope (see dashed lines in Fig. 2)
of the average velocity. For a weak rocking (small $A_{0}$) the
phenomenon of directed motion involves the surmounting of barriers.
In contrast,  for  strong rocking the averaged displacement is
related to the mean square displacement through the modified
Einstein relation, i.e.,
\begin{equation}
\kappa_2/ \mu_2(0)=k_BT_{\textmd{eff}} \;.
\end{equation}
The ballistic diffusion coefficient is $
\kappa_2=\lim_{t\to\infty}=\{\langle
x^2(t)\rangle\}_{A_0=0}/(2t^2)$, the linear mobility  $\mu_2(A_0=0)$
follows from the nonlinear mobility
$\mu_2(A_0)=\lim_{t\to\infty}\{\langle x(t)\rangle\}/(A_0 t^2)$, and
the effective temperature is
$T_{\textmd{eff}}:=T+b^2(m\{v^2(0)\}/k_B - T)$. The dissipative
acceleration of a particle of mass $m$ subjected to a constant force
$F$ yields $a=F b/m$, with $0<b<1$. This finding is therefore
intermediate  between a purely  Newton dynamics (with $b=1$) and an
ordinary  Langevin dynamics (with $b=0$). Note also that
the acceleration assumes a non-monotonic function {\it vs.} temperature.

\textit{Resume.} We have studied the nonergodic Brownian motion
occurring in so termed weak nonequilibrium baths, where the
fluctuation-dissipation theorem of the second kind still holds. The
nonequilibrium results are directly related to the ergodicity
breaking. The latter is manifested by the break-down of the FDT of
the first kind. The nonergodic behavior is either due to a vanishing
or a divergent zero-frequency spectral density of the thermal noise.
Our theory produces two limits of abnormal diffusions, namely
ballistic diffusion and a localized behavior. For ballistic
diffusion the effective friction vanishes at zero frequency  while
it assumes infinity for localization. In order to assure the ergodic
behavior of an equilibrium bath the usual conditions (i) and (ii)
must  be completed by two additional conditions: (iii) $\lim_{s\to
0}[s^{-1}\hat{\gamma} (s)]\to\infty$ and (iv) $\lim_{s\to
0}[s\hat{\gamma}(s)]=0$, where $\hat{\gamma}(s)$ is the Laplace
transform of the memory friction kernel. Yet another riveting result
is that the corresponding Brownian dynamics for a rocking Brownian
motor  exhibits a distinct, {\it accelerated}, nonstationary
velocity, rather than the constant drift which typifies the
situation with normal, Ohmic dissipation.

We are also confident that our present results will crucially impact
other quantities of thermodynamic and quantum
origin. Thus, this field is open for future studies
that in turn may reveal further surprising findings.

This work was supported by the NNSFC under 10235020 and 10475008
and the German Research Foundation DFG, Sachbeihilfe HA1517/26-1.

\end{document}